\documentstyle[12pt]{article}
 
\begin{document}

\pagestyle{myheadings}
\markboth{Zanette \& Mikhailov}{Zanette \& Mikhailov}

\title{Coherence and clustering in ensembles of neural networks}
\author{D. H.~Zanette\thanks{ 
Permanent address: Consejo Nacional de Investigaciones Cient\'{\i}ficas y
T\'ecnicas. Centro At\'omico Bariloche and Instituto Balseiro. 8400
Bariloche, Argentina.}\ \ and A. S.~Mikhailov \\
\\
Fritz-Haber-Institut der Max-Planck-Gesellschaft,\\
Faradayweg 4-6, 14195 Berlin (Dahlem), Germany}

\maketitle

\baselineskip .3 in

\vspace{.5 cm}

\centerline{\large Abstract}

Large ensembles of globally coupled chaotic neural networks undergo a 
transition to complete synchronization for high coupling intensities. 
The onset of this fully coherent behavior is preceded by a regime where
clusters of networks with identical activity are spontaneously formed. 
In these regimes of coherent collective evolution the dynamics of each 
neural network is still chaotic. These results may be relevant to
the study of systems where interaction between elements is able to give 
rise to coherent complex behavior, such as in cortex activity or in insect 
societies.

\newpage

Mutual synchronization may play a significant role in the emergence of
coherent collective behavior of different biological populations [1].
Studies of globally coupled nonlinear oscillators show the possibility of 
spontaneous transitions to a coherent regime characterized by identical
states of all elements [2-5]. Similar transitions, known as complete
stochastic synchronization, have been found for globally coupled chaotic
logistic maps [6], employed in the description of ecological dynamics [7],
and for populations of several coupled chaotic neurons [8]. However, they
deem improbable when larger and more complex systems are considered. In
contrast to this view, our study reveals that complete mutual
synchronization can easily be achieved even in ensembles whose members
represent large neural networks with complex individual dynamics. In the
emerging coherent regime, the signals generated by all networks in the
ensemble are identical. The transition is preceded by clustering, where
groups of neural networks with coherent activity patterns are spontaneously
formed. Our finding may contribute towards better understanding of coherent
activity in the cortex and in biological populations, such as swarms or
insect societies. It can find practical applications in the design of
robotic ensembles. 

Real neurons are complicated and, in theoretical studies of collective
neural activity, greatly simplified models are used [9-11]. To investigate
synchronization in ensembles of coupled neural networks, we employ the
traditional model of McCulloch and Pitts [12] where each neuron is specified
by its activity that changes in response to signals arriving from other
neurons. A network consists of a set of such elements linked through
activatory or inhibitory connections. The dynamics of such a network with
arbitrary asymmetric connections is typically characterized by an irregular
sequence of complex activity patterns.

In this Letter we consider ensembles made of $N$ identical neural networks
each consisting of $K$ neurons. The collective dynamics of an ensemble is
described by the following algorithm: At  time $t+1$, the activity $x_{k}^{i}$ 
of a neuron $k=1,...,K$ belonging to a network $i=1,...,N$ is 
\begin{equation}
x_{k}^{i}(t+1)=(1-\varepsilon )\ \Theta (h_{k}^{i})+\varepsilon \ \Theta
\left( \sum_{j=1}^{N}h_{k}^{j}\right)
\end{equation}
where $h_{k}^{i}=\sum_{l=1}^{K}J_{kl}x_{l}^{i}(t)$ is the signal arriving to
this neuron at time $t$ from all other elements of the same network, $J_{kl}$ 
are the connection weights (the same for all networks), and $\Theta (z)$ is a
sigmoidal function, such as $\Theta (z)=0$ for $z<0$ and $\Theta (z)=1$ for 
$z>0$.

The two terms in the right side of Eq. (1) have a clear interpretation. The
first of them represents the individual response of a neuron to the total
signal received from other elements in its own network. The second term
depends on the global signal obtained by summation of individual signals
received by neurons occupying the same positions in {\it all} networks of
the ensemble. The parameter $\varepsilon $ specifies the strength of global
coupling. When global coupling is absent ($\varepsilon =0$), the networks
forming the ensemble are independent. On the other hand, at $\varepsilon =1$
the first term vanishes and the states of respective neurons in all networks
must be identical since they are determined by the same global signal. For $%
0<\varepsilon <1$, the ensemble dynamics is governed by an interplay between
local coupling inside the networks and global coupling across them.

Below we show that the model exhibits, under increasing the global coupling
intensity, a spontaneous transition to a coherent collective behavior. This
transition is characterized by formation of coherent network clusters
followed by complete synchronization of all networks in the ensemble. It
takes place already at low intensities of global coupling and is observed
under an arbitrary choice of the connection weights in elementary networks.

Our analysis is based on numerical investigations. As the first step, we
have set up the connection weights between neurons in the individual
network. Each of the connection weights $J_{kl}$ between neurons has been
chosen at random with equal probability from the interval between $-1$ to 1.
The weights of forward and reverse connections were independently selected,
and therefore $J_{kl}\neq J_{lk}$. Most of the simulations were performed for
ensembles of $N=100$ identical networks, each consisting of $K=50$ neurons. 
The connection weights remained fixed within the entire series of simulations
with varying global coupling intensity. The initial conditions for all
neurons in all networks in each simulation have been randomly chosen.

Since subsequent states of all neurons in all networks are recorded, each
simulation yields a large volume of data that should be further analyzed in
order to detect coherence in the collective activity of the ensemble. An
important property is the integral time-dependent activity 
$u_{i}(t)=\sum_{k=1}^{K}x_{k}^{i}(t)$ of each network $i=1,...,N$ in the
ensemble. If global coupling is absent or very weak, the networks are
independent and, since the initial conditions are various for different
networks, their activity patterns are not correlated. Therefore, the
integral signals $u_{i}(t)$ generated by different networks in the ensemble
would be asynchronous.

Figure 1 shows typical integral signals generated by networks at higher
intensities of global coupling. The initial conditions are chosen at random
for each of the networks and therefore the integral signals of the networks
are at first not correlated. However, starting from a certain moment, some of
the networks in the ensemble begin to generate identical (up to the computer
precision) signals, indicating the onset of synchronization in the system.
When $\varepsilon =$ $0.35$ (Fig. 1A), the entire ensemble breaks eventually
down into several coherent clusters (the signals generated by networks from
the same cluster are displayed here using the same color). At a higher
intensity of global coupling ($\varepsilon =0.5$, Fig. 1B), the activity of 
{\it all} networks in the ensemble becomes eventually coherent. Quite
remarkably, the coherent signals are still very complex and apparently
chaotic.

The degree of coherence in the ensemble dynamics can be characterized by the
dispersion   of the activity patterns, defined as $D(t)=N^{-1} 
\sum_{i=1}^{N}\sum_{k=1}^{K}\left[ x_{k}^{i}(t)-\overline{x}_{k}(t)\right]
^{2}$ with $\overline{x}_{k}(t)=N^{-1}\sum_{j=1}^{N}x_{k}^{j}(t)$. Fig. 2
shows, in logarithmic scale, how this property evolves in time in a typical
simulation at a fixed intensity $\varepsilon =0.5$ of global coupling.
Within a transient period, the dispersion $D(t)$ fluctuates but remains, on
the average, approximately constant. Then, suddenly, it starts to
exponentially decrease with time and rapidly  approaches zero. When this 
occurs, full synchronization of all networks in the ensemble is achieved.

Though the dispersion serves as a good indicator of full synchronization, it
is not strongly sensitive to partial synchronization and formation of
coherent clusters in the ensemble. To analyze clustering, a different
statistical method has therefore been employed which invoved calculation of
pair distances between activity patterns of all networks. The pair distance
between the activity patterns of two networks $i$ and $j$ is defined as $%
d_{ij}=\left[ \sum_{k=1}^{K}\left( x_{k}^{i}-x_{k}^{j}\right)^2 \right]^{1/2}$. 
By counting the number of network pairs in the whole ensemble that have at
a given time the distances lying within subsequent equal intervals, a
histogram of distribution over pair distances can be constructed. Fig. 3
shows these normalized histograms for several intensities of global coupling.

When global coupling is weak ($\varepsilon =0.15$, Fig. 3A), the histogram
has a single smooth maximum at a typical distance between the activity
patterns of non-correlated networks. Increasing the coupling intensity, we
find that above a certain critical point ($\varepsilon_1 \cong
0.17$) the nature of the histogram changes. Now, some pairs of networks in
the ensemble have exactly the same activity patterns, so that the distance
between them is zero. This corresponds to the presence of a peak at $d=0$ in
Fig. 3B for $\varepsilon =0.28$. When global coupling is further increased,
the number of coherent pairs grows ($\varepsilon =0.34$, Fig. 3C).
Apparently, the coherent networks are already organized into clusters.
Indeed, several peaks are clearly seen in this histogram. The peaks are
located at pair distances between different coherent clusters. However,
besides the coherent clusters the ensemble still has a number of networks
with noncoherent activity. A slight increase of global coupling leads to the
emergence of a clear cluster organization ($\varepsilon =0.35$, Fig. 3D). In
this case, every network belongs to one of a few coherent clusters. As the
coupling intensity grows, the number of clusters gets smaller, until all
networks belong to the same coherent group. Full coherence is established in
the ensemble starting at $\varepsilon_2 \cong 0.4$ (this final regime is not
shown in Fig. 3).

We have repeated our simulations and statistical analysis for different
random choices of connection weights in the networks and have observed
basically the same sequence of changes leading to clustering and final
synchronization in all studied cases, though the respective critical
coupling intensities have been found to depend on the choice of connection
weights. Moreover, essentially the same results have been obtained when
ensembles consisting of larger networks of 100 neurons were studied and when
other sigmoidal functions $\Theta (z)$ in the algorythm (1) were employed.
This brings us to a suggestion that emergence of coherent collective
behavior might be a {\it generic} property of ensembles formed by globally
coupled neural networks.

Full mutual synchronization may play an important role in the action of
network ensembles in the brain which are responsible for generation of complex
temporal signals and/or determine motor functions. The results reported in
this Letter can also be important for studies of animal swarms and robotic 
populations. Indeed, one can imagine that each of the networks controls the 
individual behavior of a different population member. Communication between 
the members may then lead to global coupling between such networks which can,
as we have shown, result in the emergence of perfectly synchronous behaviour or
in the breakdown of the population into several coherently operating groups. 

\vspace{.5 cm}

The authors thank prof. B. Hess for enlightening discussions. D. H. Z. is
grateful to Alexander von Humboldt-Stiftung and to the Fritz Haber Institute
of the Max Planck Society for hospitality during his stay in Berlin.
Financial support from Fundaci\'{o}n Antorchas, Argentina, is acknowledged.

\newpage

\section*{Figure captions}

\begin{itemize}
\item[Fig.~1:]  Time-dependent integral activity of ten selected networks in an
ensemble of $N=100$ for different intensities of global coupling, corresponding 
to (A) clustering ($\varepsilon =0.35$) and (B) complete synchronization 
($\varepsilon =0.5$). Synchronous signals are indicated by the same colors.

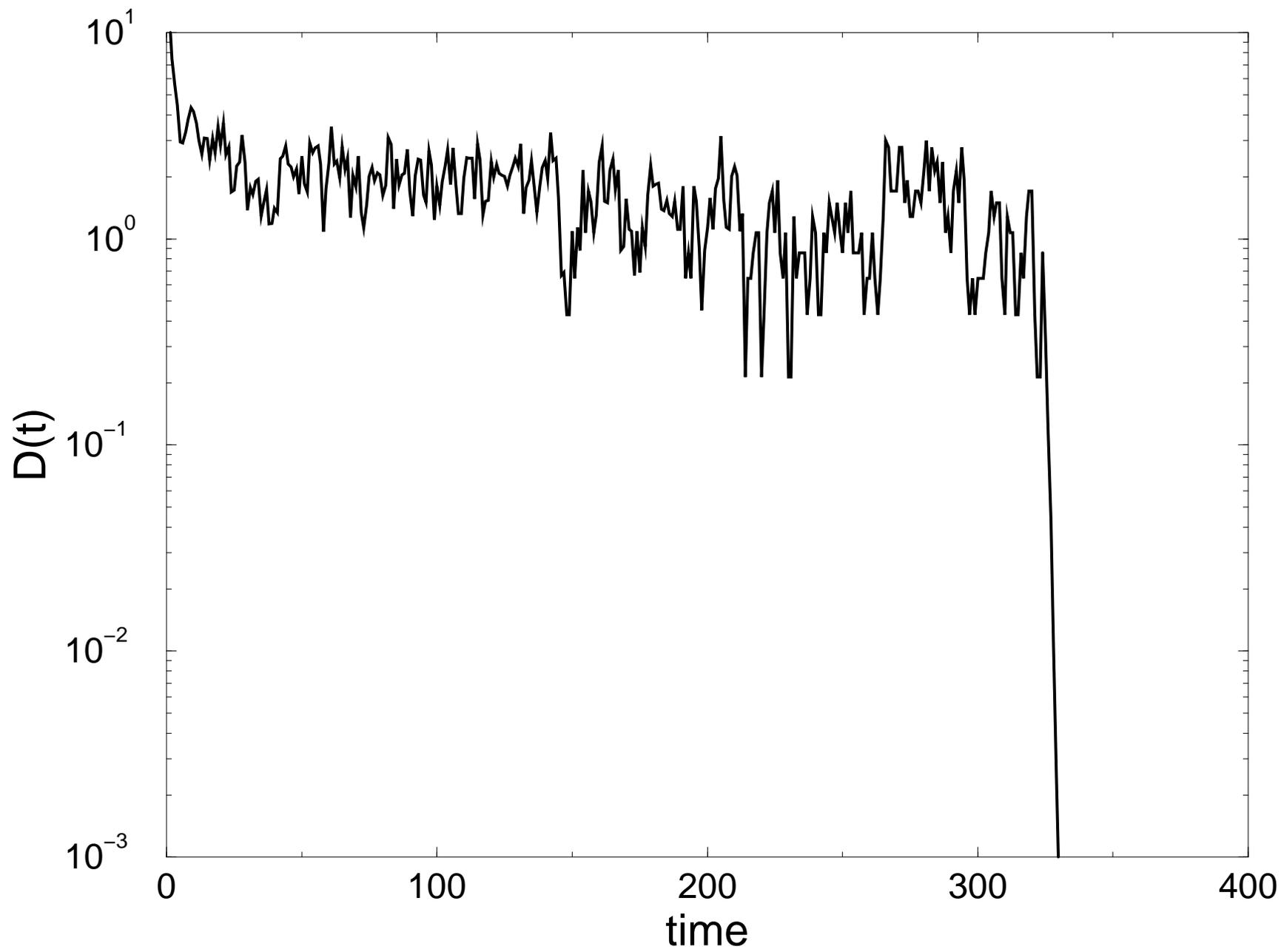
\item[Fig.~2:]  Dispersion of the activity patterns of all networks in the
ensemble as function of time under  synchronization conditions ($\varepsilon
=0.5$).

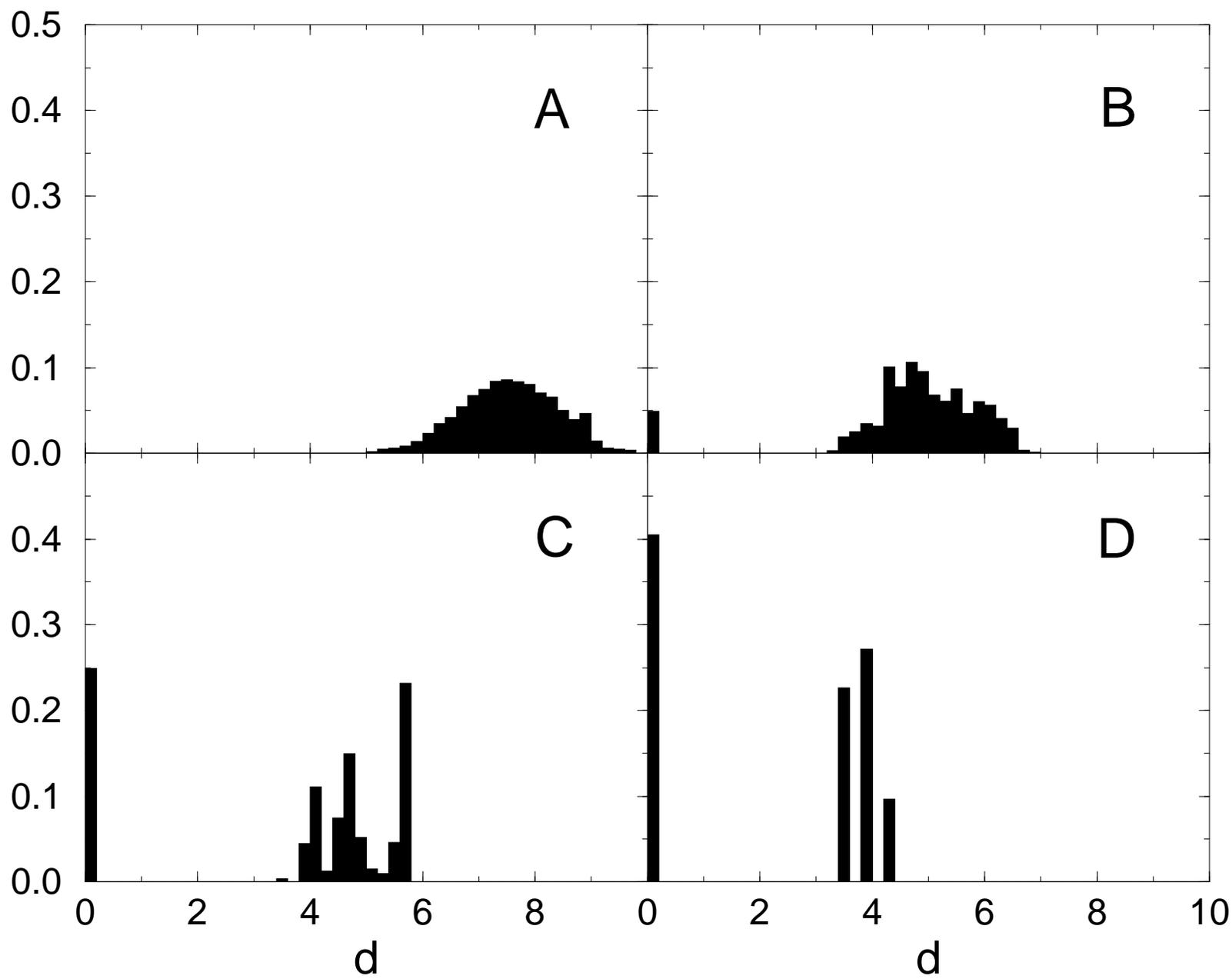
\item[Fig.~3:]  Histograms of distributions over pair distances $d$ between
activity patterns of all networks in the ensemble for different intensities
of global coupling corresponding to (A) asynchronous regime, $\varepsilon
=0.15$, (B) onset of synchronization, $\varepsilon =0.28$, (C) partial
clustering, $\varepsilon =0.34$, and (D) full clustering, $\varepsilon =0.35$.
\end{itemize}

\end{document}